\documentclass[11pt,english]{article}
\pdfoutput=1
\usepackage[T1]{fontenc}
\usepackage[latin9]{inputenc}
\usepackage{geometry}
\usepackage{units}
\usepackage{amssymb}
\usepackage{graphicx}
\usepackage{setspace}
\makeatletter

\newcommand{\lyxaddress}[1]{
\par {\raggedright #1
\vspace{1.4em}
\noindent\par}
}

\usepackage{fourier}
\usepackage{babel}
\usepackage{amssymb}
\usepackage[sort&compress,numbers]{natbib}
\usepackage{caption}
\DeclareCaptionLabelSeparator{bar}{ | }
\DeclareCaptionFormat{tight}{\setlength{\baselineskip}{3.354mm}#1#2#3\par}
\small
\date{}

\makeatother

\usepackage{babel}
\begin{document}

\title{Demonstration of Entanglement of Electrostatically Coupled Singlet-Triplet
Qubits}

\author{M. D. Shulman\textsuperscript{1$\dagger$}, O. E. Dial\textsuperscript{1$\dagger$},
S. P. Harvey\textsuperscript{1}, H. Bluhm\textsuperscript{1{*}},
V. Umansky\textsuperscript{2}, and A. Yacoby\textsuperscript{1}}

\maketitle

\lyxaddress{\textsuperscript{1}Department of Physics, Harvard University, Cambridge,
MA, 02138, USA}

\lyxaddress{\textsuperscript{2}Braun Center for Submicron Research, Department
of Condensed Matter Physics, Weizmann Institude of Science, Rehovot
76100 Israel}

\lyxaddress{\textsuperscript{{*}}Current Address: 2nd Institute of Physics C,
RWTH Aachen University, 52074 Aachen, Germany}

\lyxaddress{\textsuperscript{$\dagger$}These authors contributed equally to
this work. }
\begin{abstract}
\textbf{Quantum computers have the potential to solve certain interesting
problems significantly faster than classical computers. To exploit
the power of a quantum computation it is necessary to perform inter-qubit
operations and generate entangled states. Spin qubits are a promising
candidate for implementing a quantum processor due to their potential
for scalability and miniaturization. However, their weak interactions
with the environment, which leads to their long coherence times, makes
inter-qubit operations challenging. We perform a controlled two-qubit
operation between singlet-triplet qubits using a dynamically decoupled
sequence that maintains the two-qubit coupling while decoupling each
qubit from its fluctuating environment. Using state tomography we
measure the full density matrix of the system and determine the concurrence
and the fidelity of the generated state, providing proof of entanglement. }
\end{abstract}
Singlet-triplet ($S$-$T_{0}$) qubits, a particular realization of
spin qubits\cite{Loss1998,Koppens2006,Pioro2008,Nowack2007,Pioro2007,Churchill2009,Nadj-Perge2010},
store quantum information in the joint spin state of two electrons\cite{Levy2002,Petta2005,Taylor2007}.\textbf{
}The basis states for the $S$-$T_{0}$ qubit can be constructed from
the eigenstates of a single electron spin, $|\uparrow\rangle$ and
$|\downarrow\rangle$. We choose $|S\rangle=\frac{1}{\sqrt{2}}\left(|\uparrow\downarrow\rangle-|\downarrow\uparrow\rangle\right)$
and $|T_{0}\rangle=\frac{1}{\sqrt{2}}\left(|\uparrow\downarrow\rangle+|\downarrow\uparrow\rangle\right)$
for the logical subspace of the $S$-$T_{0}$ qubit because these
states are insensitive to uniform fluctuations in the magnetic field.
The qubit can then be described as a two level system with a representation
on a Bloch sphere shown in Fig. 1a. Universal quantum control is achieved
using two physically distinct operations that drive rotations around
the x and z-axes of the Bloch sphere \cite{Foletti2009}. Rotations
around the z-axis of the Bloch sphere are driven by the exchange splitting,
$J$, between $|S\rangle$ and $|T_{0}\rangle$, and rotations around
the x-axis are driven by a magnetic field gradient, $\Delta B_{z}$
between the electrons. 

We implement the $S$-$T_{0}$ qubit by confining two electrons to
a double quantum dot (QD) in a two dimensional electron gas (2DEG)
located 91nm below the surface of a GaAs-AlGaAs heterostructure. We
deposit local top gates using standard electron beam lithography techniques
in order to locally deplete the 2DEG and form the QDs. We operate
between the states (0,2) and (1,1) where ($n_{L}$,$n_{R}$) describes
the state with $n_{L}$($n_{R}$) electrons in the left (right) QD.
The $|S\rangle$ and $|T_{0}\rangle$ states, the logical subspace
for the qubit, are isolated by applying an external magnetic field
of $\mathbf{B}=$700mT in the plane of the device such that the Zeeman
splitting makes $T_{+}=|\uparrow\uparrow\rangle$, and $T_{-}=|\downarrow\downarrow\rangle$
energetically inaccessible. The exchange splitting, $J$, is a function
of the difference in energy, $\epsilon$, between the levels of the
left and right QDs. Pulsed DC electric fields rapidly change $\epsilon$,
allowing us to switch $J$ on, which drives rotations around the z-axis.
When $J$ is off the qubit precesses around the x-axis due to a fixed
$\Delta B_{z}$, which is stabilized to $\nicefrac{\Delta B_{z}}{2\pi}=$30MHz
by operating the qubit as a feedback loop between interations of the
experiment\cite{Bluhm2009}. Dephasing of the qubit rotations reflects
fluctuations in the magnitude of the two control axes, $J$ and $\Delta B$,
caused by electrical noise and variation in the magnetic field gradient,
respectively. The qubit is rapidly (<50ns) initialized in $|S\rangle$
by exchanging an electron with the nearby Fermi sea of the leads of
the QD in a region of (0,2) where only $|S\rangle$ is accessible,
and the qubit state is read out using standard Pauli blockade techniques,
where $\epsilon$ is quickly tuned to the regime where S occupies
(0,2) and $T{}_{0}$ occupies (1,1), allowing the qubit state to be
determined by the proximal charge sensor. The charge state of the
qubit is rapidly read ($\sim1\mu s$) using standard RF-techniques
\cite{Reilly2007,Barthel2009} on an adjacent sensing QD.

\begin{doublespace}
In order to make use of the power of quantum information processing
it is necessary to perform two qubit operations in which the state
of one qubit is conditioned on the state of the other\cite{Nielsen2000}.
To investigate two-qubit operations we fabricate two adjacent $S$-$T_{0}$
qubits such that they are capacitively coupled, but tunneling between
them is suppressed (Fig. 1b). A charge sensing QD next to each qubit
allows for simultaneous and independent projective measurement of
each qubit (supplement). We use the electrostatic coupling between
the qubits to generate the two-qubit operation\cite{Taylor2005}.
When $J$ is nonzero, the $S$ and $T_{0}$ states have different
charge configurations in the two QDs due to the Pauli exclusion principle
(Fig. 1c). This charge difference, which is a function of $\epsilon$,
causes the $|S\rangle$ and $|T_{0}\rangle$ states in one qubit to
impose different electric fields on the other qubit. Since $J$ is
a function of the electric field, the change imposed by the first
qubit causes a shift in the precession frequency of the second qubit.
In this way the state of the second qubit may be conditioned on the
state of the first qubit. More precisely, when a single qubit evolves
under exchange, there exists a state-dependent dipole moment, $\vec{d}$,
between $|S\rangle$ and $|T{}_{0}\rangle$ resulting from their difference
in charge occupation of the QDs. Therefore, when simultaneously evolving
both qubits under exchange, they experience a capacitively mediated,
dipole-dipole coupling that can generate an entangled state. The two-qubit
Hamiltonian is therefore given by:
\begin{equation}
\mathcal{H}_{2-qubit}=\frac{\hbar}{2}\left(J_{1}(\sigma_{z}\otimes I)+J_{2}(I\otimes\sigma_{z})+\frac{J_{12}}{2}\left(\left(\sigma_{z}-I\right)\otimes\left(\sigma_{z}-I\right)\right)+\Delta B_{z,1}(\sigma_{x}\otimes I)+\Delta B_{z,2}(I\otimes\sigma_{x})\right)\label{eq:H_2qubit}
\end{equation}
 where $\sigma_{x,y,z}$ are the Pauli matrices, $I$ is the identity
operator, $\Delta B_{z,i}$ and $J_{i}$ are the magnetic field gradients
and the exchange splittings (i=1,2 for the two qubits), and $J_{12}$
is the two-qubit coupling, which is proportional to the product of
the dipole moments in each qubit. For a two level system with constant
tunnel coupling, the dipole moment scales as $\vec{d}_{i}\propto\frac{\partial J_{i}}{\partial\epsilon_{i}}$.
Empirically, we find that for experimentally relevant values of $J_{i}$,
$\frac{\partial J_{i}}{\partial\epsilon_{i}}\propto J_{i}(\epsilon)$,
so that $J_{12}\propto J_{1}J_{2}$. As with the single qubit operations,
this two-qubit operation requires only pulsed DC electric fields. 
\end{doublespace}

In principle, evolving both qubits under exchange produces an entangling
gate. However, the time to produce this maximally entangled state
exceeds the inhomogeneously broadened coherence times of each individual
qubit, rendering this simple implementation of the two-qubit gate
ineffective. To mitigate this we use a dynamically decoupled entangling
sequence\cite{Viola1999C,Leibried2003}(Fig. 1d). In this sequence,
each qubit is prepared in $|S\rangle$ and is then rotated by $\frac{\pi}{2}$
around the x-axis ($J_{i}=$0, $\Delta B_{z,i}/2\pi\approx$30MHz)
to prepare a state in the x-y plane. The two qubits are subsequently
both evolved under a large exchange splitting $\left(J_{1}/2\pi\approx280MHZ,\ J_{2}/2\pi\approx320MHz\gg\Delta B_{z}\right)$
for a time $\frac{\tau}{2}$, during which the qubits begin to entangle
and disentangle. A $\pi$-pulse around the x-axis ($\Delta B_{z}$)
is then applied simultaneously to both qubits, after which the qubits
are again allowed to exchange for a time $\frac{\tau}{2}$. This Hahn
echo-like sequence\cite{Hahn1950} removes the dephasing effect of
noise that is low frequency compared to $1/\tau$, while the $\pi$-pulses
preserve the sign of the two-qubit interaction. The resulting operation
produces a CPHASE gate, which, in a basis of $\left\{ |SS\rangle,|T_{0}S\rangle,|ST_{0}\rangle,|T_{0}T_{0}\rangle\right\} $,
is an operation described by a matrix with $e^{-i\theta/2}$,1,1,$e^{-i\theta/2}$
on the diagonals. For $\tau=\tau_{ent}=\frac{\pi}{2J_{12}}$ ($\theta=\pi$)
the resulting state is a maximally entangled generalized Bell state
$|\Psi_{ent}\rangle=e^{i\pi(I\otimes\sigma_{y}+\sigma_{y}\otimes I)/8}|\Psi_{-}\rangle$,
which differs from the Bell state $|\Psi_{-}\rangle=\frac{1}{\sqrt{2}}\left(|SS\rangle-|T_{0}T_{0}\rangle\right)$
by single qubit rotations.

\begin{doublespace}
In order to characterize our two-qubit gate and verify that we produce
an entangled state we perform two-qubit state tomography and extract
the density matrix and appropriate entanglement measures. The tomographic
procedure is carefully calibrated with minimal assumptions in order
to avoid adding spurious correlations to the data that may artificially
increase the measured degree of entanglement (supplement). We choose
the Pauli set representation of the density matrix\cite{James_tomo_2001,Nielsen2000,Chow2010},
where we measure and plot the 16 two-qubit correlators $\langle ij\rangle=\langle\sigma_{i}\sigma_{j}\rangle$
where $\sigma_{i}$ are the Pauli matrices and $i,j\in\left\{ I,X,Y,Z\right\} $.
As a first measure of entanglement, we evaluate the concurrence\cite{Hill_and_Wooters_1997}
(Fig. 2a), $C\left(\rho\right)=\max\left\{ 0,\,\lambda_{4}-\lambda_{3}-\lambda_{2}-\lambda_{1}\right\} $
for different $\tau$, where $\rho$ is the experimentally measured
density matrix, and $\lambda_{i}$ are the eigenvalues, sorted from
largest to smallest, of the matrix $R=\sqrt{\sqrt{\rho}\tilde{\rho}\sqrt{\rho}}$,
and $\tilde{\rho}=(\sigma_{y}\otimes\sigma_{y})\rho^{*}(\sigma_{y}\otimes\sigma_{y})$,
and $\rho^{*}$ is the complex conjugate of $\rho$. A positive value
of the concurrence is a necessary and sufficient condition for demonstration
of entanglement \cite{Hill_and_Wooters_1997}. For $\tau=$140ns we
extract a maximum concurrence of 0.44. 
\end{doublespace}

While a positive value of the concurrence is a definitive proof of
entanglement, it alone does not verify that the two-qubit operation
produces the intended entangled state. In order to better characterize
the generated quantum state, we evaluate another measure of entanglement,
the Bell state fidelity, $F\equiv\langle\Psi_{ent}|\rho|\Psi_{ent}\rangle$.
This may be interpreted as the probability of measuring our two-qubit
state in desired $|\Psi_{ent}\rangle$. Additionally, for all non-entangled
states one can show that $F\leq0.5$ \cite{SacketFidelity2000,Bennet1996}.
In terms of the Pauli basis, the Bell state fidelity takes the simple
form $F=\frac{1}{4}\vec{P}_{ent}\cdot\vec{P}_{experiment}$ where
$\vec{P}_{ent}$ and $\vec{P}_{experiment}$ are the Pauli sets of
a pure target Bell state and of the experimentally measured state,
respectively. For our target state $|\Psi_{ent}\rangle$, the resulting
Pauli set is given by $\langle XZ\rangle=\langle ZX\rangle=\langle YY\rangle=$1,
with all other elements equal to zero (Fig 3a).

In an idealized, dephasing-free version of the experiment, as $\tau$
increases and the qubits entangle and disentangle, we expect the nonzero
elements of the Pauli set for the resulting state to be 
\begin{equation}
\langle YI\rangle=\langle IY\rangle=\cos\left(J_{12}\tau\right),\;\langle XZ\rangle=\langle ZX\rangle=\sin\left(J_{12}\tau\right),\;\langle YY\rangle=1\label{eq: ideal pauli set}
\end{equation}
Dephasing due to electrical noise causes the amplitudes of the Pauli
set to decay. However, the two-qubit Hamiltonian (equation(\ref{eq:H_2qubit}))
includes rapid single-qubit rotations around the S-$T_{0}$ axis ($J_{1}$,$J_{2}\gg$$J_{12}/2\pi\approx1MHz$)
that change with $\tau$ due to imperfect pulse rise times in the
experiment. These add additional single-qubit rotations around the
$S$-$T_{0}$ axis of each qubit, which are not accounted for in equation
(\ref{eq: ideal pauli set}). We determine the angle of the single
qubit rotations by performing a least-squares fit of the experimental
data to modified form of equation (\ref{eq: ideal pauli set}) that
accounts for these rotations and dephasing. The decays due to dephasing
are fit by calculating $\rho\left(t\right)$ in the presence of noise
on $J_{1}$ and $J_{2}$, which leads to decay of certain terms in
the density matrix \cite{Yu2006,Cywinski2008B}. For the present case
where $J_{12}\ll J_{1},\, J_{2}$ , we neglect the two-qubit dephasing,
which is smaller than single-qubit dephasings by a factor of $\frac{J_{1}}{J_{12}}$,
$\frac{J_{2}}{J_{12}}\approx$300, and we extract a separate dephasing
time for each individual qubit. We remove the single-qubit rotations
numerically in order to simplify the presentation of the data (Fig.
3e). The extracted angles exhibit a smooth monotonic behavior which
is consistent with their underlying origin (see supplemental information).

\begin{doublespace}
In the absence of dephasing we would expect the Bell state fidelity
to oscillate between 0.5 for an unentangled state and 1 for an entangled
state as a function of $\tau$. This oscillation is caused by the
phase accumulated by a CPHASE gate between the two qubits. However,
the qubits dephase as the state becomes increasingly mixed, and this
oscillation decays to 0.25. Indeed, this behavior is observed (Fig.
2b): for very short $\tau$ there is very little dephasing present,
and the qubits are not entangled. As $\tau$ increases the Bell state
fidelity increases as the qubits entangle, reaching a maximum value
of 0.72 at $\tau=$140ns. As $\tau$ is increased further, we continue
to see oscillations in the Bell state fidelity , though due to dephasing,
they do not again rise above 0.5.

Fig. 2c shows these oscillations in Bell state fidelity as a function
of $\tau$ for several different values of $J$ as $\epsilon$ is
changed symmetrically in the two qubits. We see that as the value
of $J$ increases in the two qubits, the time required to produce
a maximally entangled state, $\tau_{ent}$, decreases, but the maximum
attainable fidelity is approximately constant. This is consistent
with the theory that $J_{12}\propto\frac{\partial J_{1}}{\partial\epsilon_{1}}\cdot\frac{\partial J_{2}}{\partial\epsilon_{2}}\propto J_{1}\cdot J_{2}$. 

To further understand the evolution of the quantum state, we focus
on one value of $J$ and compare the measured Pauli set to that expected
from single-qubit dephasing rates and $J_{12}$ (see supplemental
information). Fig. 3a shows the Pauli set for the measured and expected
quantum states for $\tau=$40ns, which shows three large bars in the
$\langle YI\rangle$,$\langle IY\rangle$, and $\langle YY\rangle$
components of the Pauli set. This is a nearly unentangled state. At
$\tau=$140ns, we see weight in the in the $\langle XZ\rangle$, $\langle ZX\rangle$,
$\langle YY\rangle$ components of the Pauli set (Fig. 3b), and we
extract a Bell state fidelity of 0.72, which demonstrates the production
of an entangled state. For $\tau=\tau_{ent}=\frac{\pi}{2J_{12}}$=
160ns (Fig. 3c) we see a similar state to $\tau=$140ns, but with
less weight in the single qubit components of the Pauli set. This
state corresponds to the intended CPHASE of $\pi$, though it has
a slightly lower fidelity than the state at $\tau=$140ns due to additional
decoherence. Finally, at $\tau=\frac{\pi}{J_{12}}=$320ns (Fig. 3d),
where we expect the state to be unentangled, we again see large weight
in the $\langle YI\rangle$, $\langle IY\rangle$, and $\langle YY\rangle$
components of the Pauli set, though the bars are shorter than the
Pauli set for $\tau=$40ns, due to dephasing of the qubits. We plot
the entire Pauli set as a function of time (Fig. 3e), which clearly
shows the predicted oscillation (equation(\ref{eq: ideal pauli set}))
between $\langle YI\rangle$,$\langle IY\rangle$ and $\langle XZ\rangle$,
$\langle ZX\rangle$, with decays due to decoherence. 

The two-qubit gate that we have demonstrated is an important step
toward establishing a scalable architecture for quantum information
processing in $S$-$T_{0}$ qubits. State fidelity is lost to dephasing
from electrical noise, and decreasing the ratio $\frac{\tau_{ent}}{T_{2}^{echo}}$,
where $T_{2}^{echo}$ is the single-qubit coherence time with an echo
pulse, is therefore paramount to generating high-fidelity Bell states.
Large improvements can be made by introducing an electrostatic coupler
between the two qubits \cite{Trifunovic2012} in order to increase
the two-qubit coupling ($J_{12}$) and reduce $\tau_{ent}$. We estimate
that in the absence of other losses, if an electrostatic coupler were
used, a Bell state with fidelity exceeding 90\% could be produced.
Additional improvements can be made by studying and mitigating the
origins of charge noise to increase $T_{2}^{echo}.$ This would allow
future tests of complex quantum operations including quantum algorithms
and quantum error correction. The generation of entangled also states
opens the possibility of studying the complex dynamics of the nuclear
environment, which is a fundamental, quantum, many body problem.
\end{doublespace}

\section*{Acknowledgments}

This work is supported through the NSA, \textquotedblleft{}Precision
Quantum Control and Error-Suppressing Quantum Firmware for Robust
Quantum Computing\textquotedblright{} and IARPA \textquotedblleft{}Multi-Qubit
Coherent Operations (MQCO) Program.\textquotedblright{} This work
was partially supported by the US Army Research Office under Contract
Number W911NF-11-1-0068. This work was performed in part at the Center
for Nanoscale Systems (CNS), a member of the National Nanotechnology
Infrastructure Network (NNIN), which is supported by the National
Science Foundation under NSF award no. ECS-0335765. CNS is part of
Harvard University.

\section*{Author Contributions}

V.U. prepared the crystal, M.D.S. fabricated the sample, M.D.S., O.E.D.,
H.B., S.P.H., A.Y. carried out the experiment, analyzed the data,
and wrote the paper.

\section*{Additional Information}

The authors declare no competing financial interests. Supplementary
information accompanies this paper etc etc. Correspondence should
be directed to AY at yacoby@physics.harvard.edu 

\bibliographystyle{unsrt}
\bibliography{SF_plus_more}

\pagebreak{}

\begin{figure}
\includegraphics[scale=0.5]{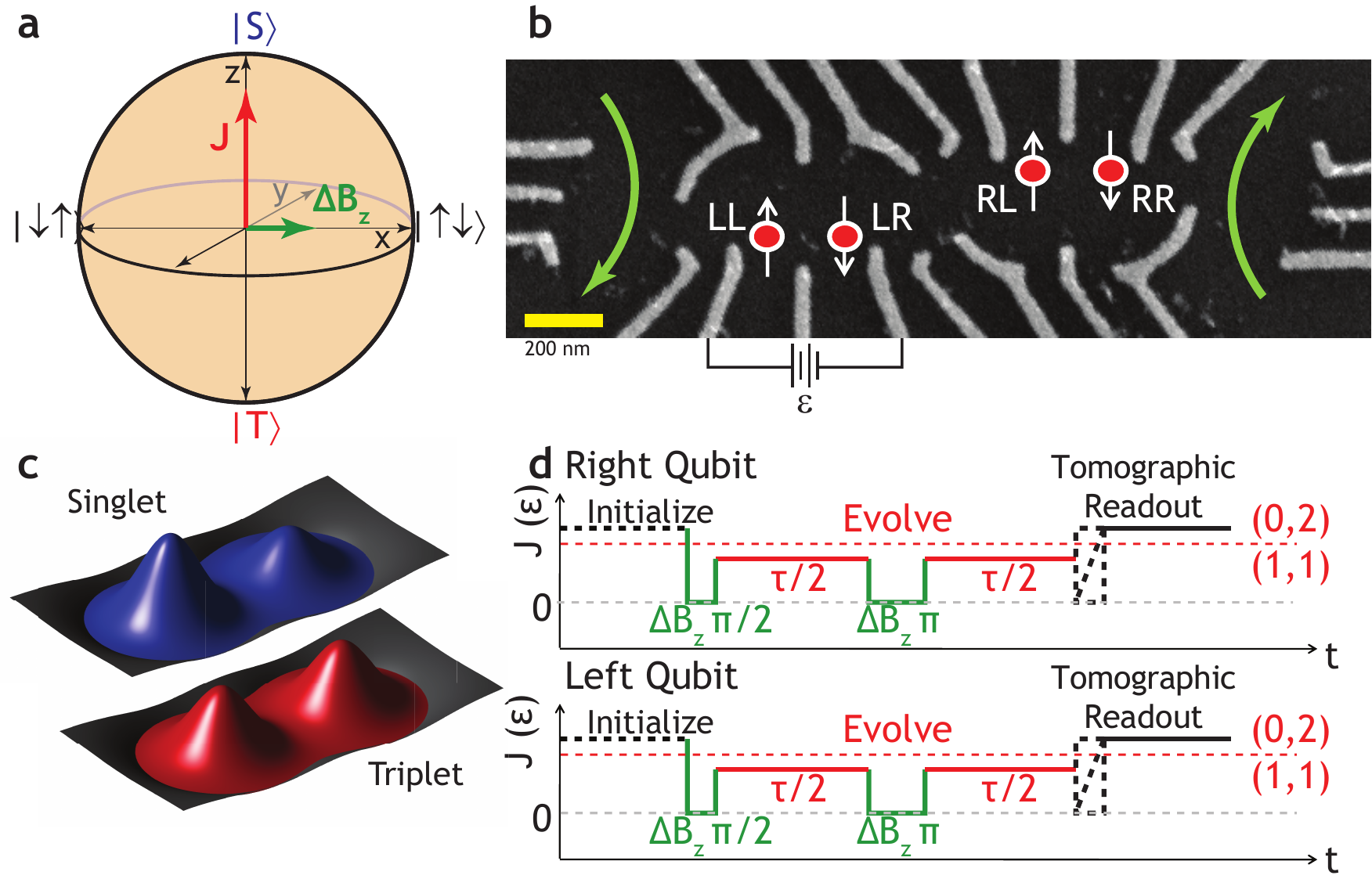}\caption{\textbf{Two-qubit coupling scheme. a}, A Bloch sphere can be used
to describe the states of the effective two-level system defined by
the singlet and triplet states of the qubit, with the z-axis along
the $S$-$T_{0}$ axis and the x-axis along the $\left|\uparrow\downarrow\right\rangle $/$\left|\downarrow\uparrow\right\rangle $
axes. \textbf{b}, An SEM image of the top of the device used shows
gates used to define the $S$-$T_{0}$ qubits (white), dedicated ns
control leads, the approximate locations of the electrons in the two
qubits (red), and current paths for the sensing dots (green arrows).
The left qubit uses the LR and LL electrons, while the right qubit
uses the RL and RR electrons. \textbf{c}, A schematic of the electronic
charge configurations for the $|S\rangle$ (blue) and the $|T_{0}\rangle$
(red) at non-zero $J$. This difference in charge configuration is
the basis for the electrostatic coupling between the qubits\textbf{
d}, The pulse sequence used to entangle the qubits: initialize each
qubit in a $|S\rangle$, perform a $\pi/2$ rotation around the x-axis,
allow the qubits to evolve under exchange for a time $\tau/2$, perform
a $\pi$-rotation around the x-axis, thereby decoupling the qubits
from the environment but not each-other, evolve under exchange for
$\tau/2$, and perform state tomography to determine the resulting
density matrix (see supplemental information.)}
 
\end{figure}

\pagebreak{}

\begin{figure}
\includegraphics[scale=0.5]{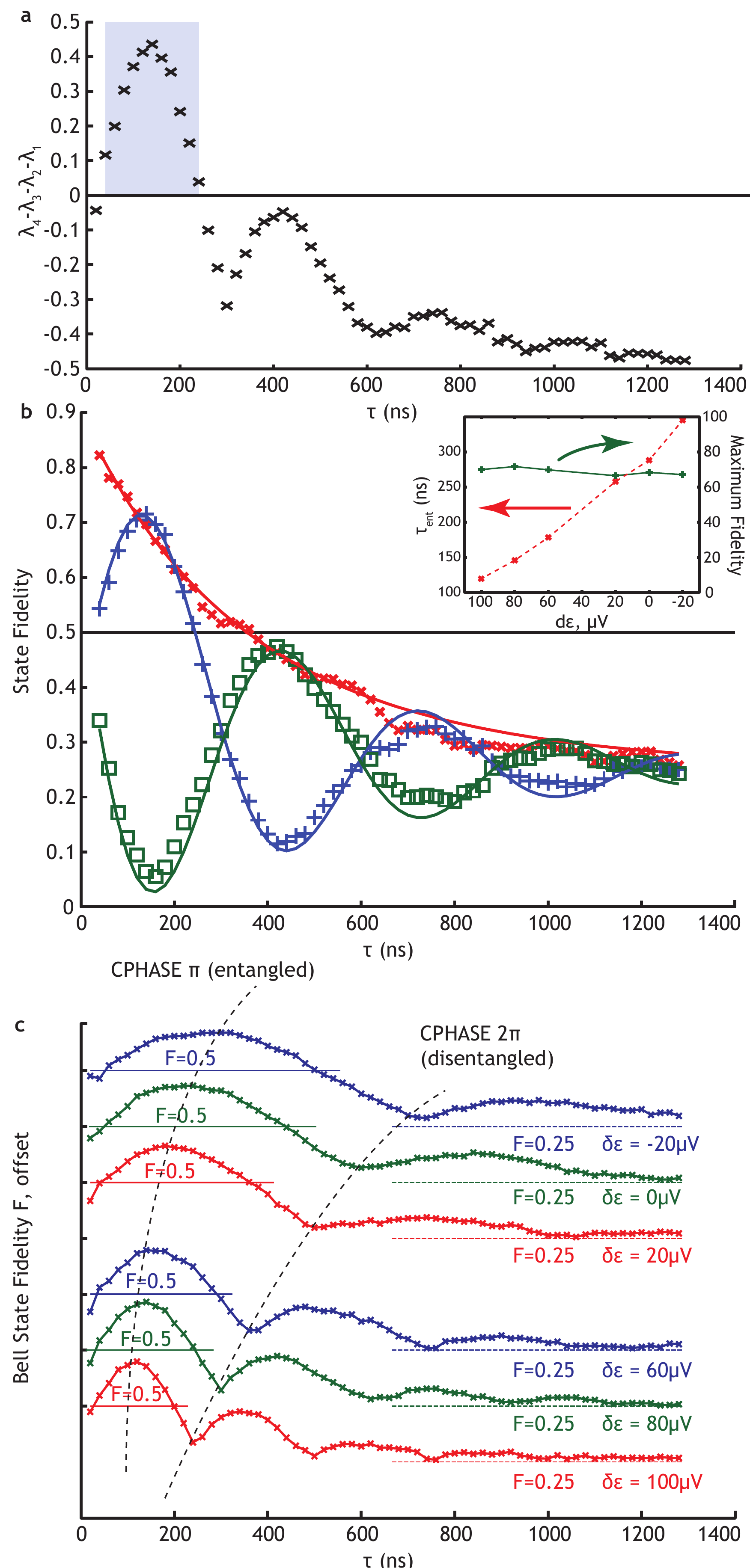}\caption{\textbf{Proof of entanglement: concurrence and state fidelity. a},
A plot of the difference of the sorted eigenvalues of the matrix R,
which for positive values is equal to the concurrence $c\left(\rho\right)$.
States with a concurrence greater than zero (shaded region) are necessarily
entangled. \textbf{b}, The fidelity with which the measured state
approximates the target $|\Psi_{ent}\rangle$ (blue), and $e^{i\pi\left(\sigma_{y}\otimes I+I\otimes\sigma_{y}\right)/4}|\Psi_{ent}\rangle$
(green), which differs from $|\Psi_{ent}\rangle$ by single qubit
rotations and is the expected state for $\tau=\nicefrac{3\pi}{2J_{12}}$.
The fidelity with which the measured state approximates a dephasing-free
model of the entangling operation (red) shows smooth decay due to
decoherence. The solid lines are fits to the data. Inset: The time
to produce a maximally entangled state as a function of the change
in $\epsilon$ (and therefore $J$) in the two qubits. As $J$ increases
$\tau_{ent}$ decreases (red), but the maximum attainable fidelity
(green) is approximately constant. Arrows indicate which y-axis is
to be used. \textbf{c}, The Bell state fidelity as a function of time
for different values of $J$ (offset) with guides to show where the
fidelity exceeds $\nicefrac{1}{2}$ for each curve. As $J$ increases
in the two qubits the time to produce an entangled state, $\tau_{ent}$,
decreases.}
\end{figure}

\pagebreak{}

\begin{figure}
\includegraphics[scale=0.5]{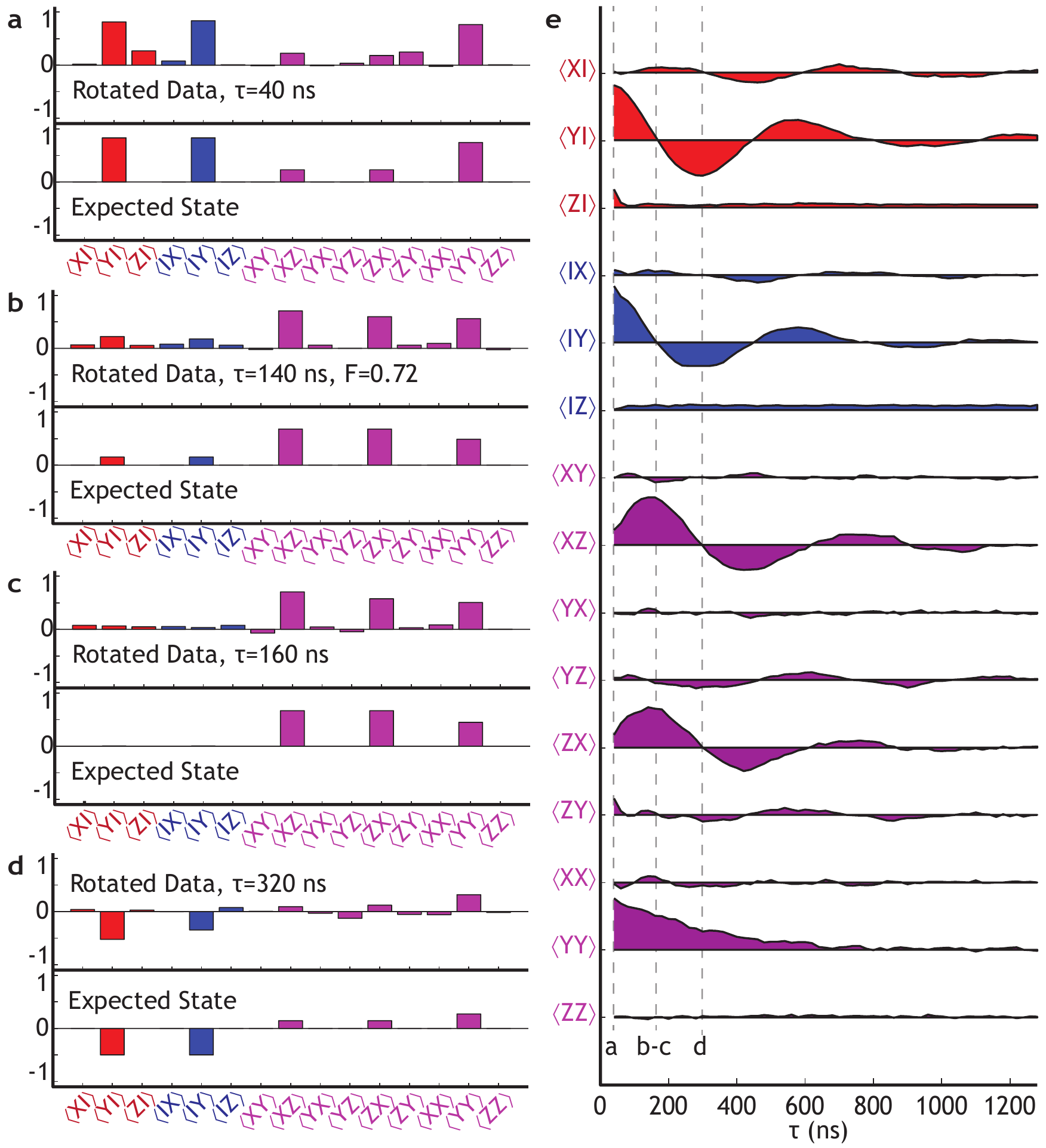}\caption{\textbf{Pauli set representation. a, }The\textbf{ }elements of the
Pauli set for the measured density matrix and the state expected from
the entangling and dephasing rates for small $\tau$ (unentangled).\textbf{
b}, The Pauli set of the measured and expected states for $\tau=$140ns,
which produces a maximum Bell state fidelity of 0.72. \textbf{c},
The Pauli set of the measured and expected states for $\tau=\tau_{ent}=$160ns,
which is a CPHASE of $\pi$ but does not the highest fidelity due
to dephasing. \textbf{d}, The Pauli set for the measured and expected
states for $\tau=$320ns, which is an unentangled state. \textbf{e},
The full measured Pauli set as a function of $\tau$, which shows
the expected behavior for a CPHASE gate. The the y-axes of adjacent
elements in the Pauli set are offset by 1. }
 
\end{figure}

\end{document}

\title{Supplementary Information for ``Demonstration of Entanglement of
Electrostatically Coupled Singlet-Triplet Qubits''}

\author{M. D. Shulman\textsuperscript{1}, O. E. Dial\textsuperscript{1},
S. P. Harvey\textsuperscript{1}, H. Bluhm\textsuperscript{1{*}},
V. Umansky\textsuperscript{2}, and A. Yacoby\textsuperscript{1}\\
{\small \textsuperscript{1}Department of Physics, Harvard University,
Cambridge, MA, 02138, USA}\\
{\small \textsuperscript{2}Braun Center for Submicron Research, Department
of Condensed Matter Physics, Weizmann Institute of Science, Rehovot
76100 Israel}\\
{\small \textsuperscript{{*}}Current Address: 2nd Institute of Physics
C, RWTH Aachen University, 52074 Aachen, Germany}\\
}

\maketitle
\textbf{\Large Calibration of RF Sensor Response}{\Large \par}

In order to quantitatively interpret sensor values for state tomography,
it is important to precisely determine the RF sensor response that
corresponds to a $|S\rangle$ or a $|T_{0}\rangle$ state. Because
the state preparation is imperfect, it is in general difficult to
accurately measure these values experimentally. To provide exact calibrations
for $|S\rangle$ and $|T_{0}\rangle$, we exploit the fact that our
sensor is capable of single shot readout. Histograms of sensor values
for typical measurements yield a double-peaked curve- one peak corresponds
to $|T_{0}\rangle$ and one to $|S\rangle$ (Suppl. Fig 1a). In order
to calibrate the sensor we first measure $T_{1}$ at the measurement
point by preparing a state that is majority $|T_{0}\rangle$ (done
with a $\pi$-pulse around the x-axis) and fitting the sensor signal
to a decaying exponential function of time elapsed before measurement
(Suppl Fig. 1a). We note that the measured value of $T_{1}$ is a
strong function of the power of the RF excitation used to read the
conductance of the sensing QD. With prior knowledge of $T_{1}$, we
use a procedure similar to that described in Barthel et. al (ref.
14) to optimize the measurement time given our signal to noise ratios
and $T_{1}$. This process is repeated several times per day to check
for drift. We recalibrate the sensor signals that correspond to $|S\rangle$
and $|T_{0}\rangle$ for each dataset (typically 10 minutes of acquisition
time). For each set, we prepare a histogram of all observed sensor
values. The presence of several reference measurements in each dataset
guarantees that there will be a significant fraction of both $|S\rangle$
and $|T_{0}\rangle$. We then fit this double peaked curve to an analytic
expression corresponding to a weighted sum of two Gaussians with some
filling in due to $T_{1}$ decay during measurement (Suppl. Fig. 1b,
purple line) as in ref. 14. From this, we extract the expected sensor
distributions for $|S\rangle$ and $|T_{0}\rangle$ (blue and red
lines in Suppl. Fig. 1b, respectively), as well as the fractions of
$|S\rangle$ and $|T_{0}\rangle$ present. The centers of the two
distributions correspond to the sensor signals that will be measured
for pure $|S\rangle$ and pure $|T_{0}\rangle$, and using these values
we can accurately scale the tomography data. We note that this procedure
is insensitive to the percentages of $|S\rangle$ and $|T_{0}\rangle$.
In our state tomography only expectation values are needed, so the
single-shot capability of our readout is not necessary beyond this
calibration. Nonetheless, we note that for the data presented, we
measured readout fidelities of 97\% and 98\% for the left and right
qubits, respectively.

\begin{figure}
\includegraphics[scale=0.5]{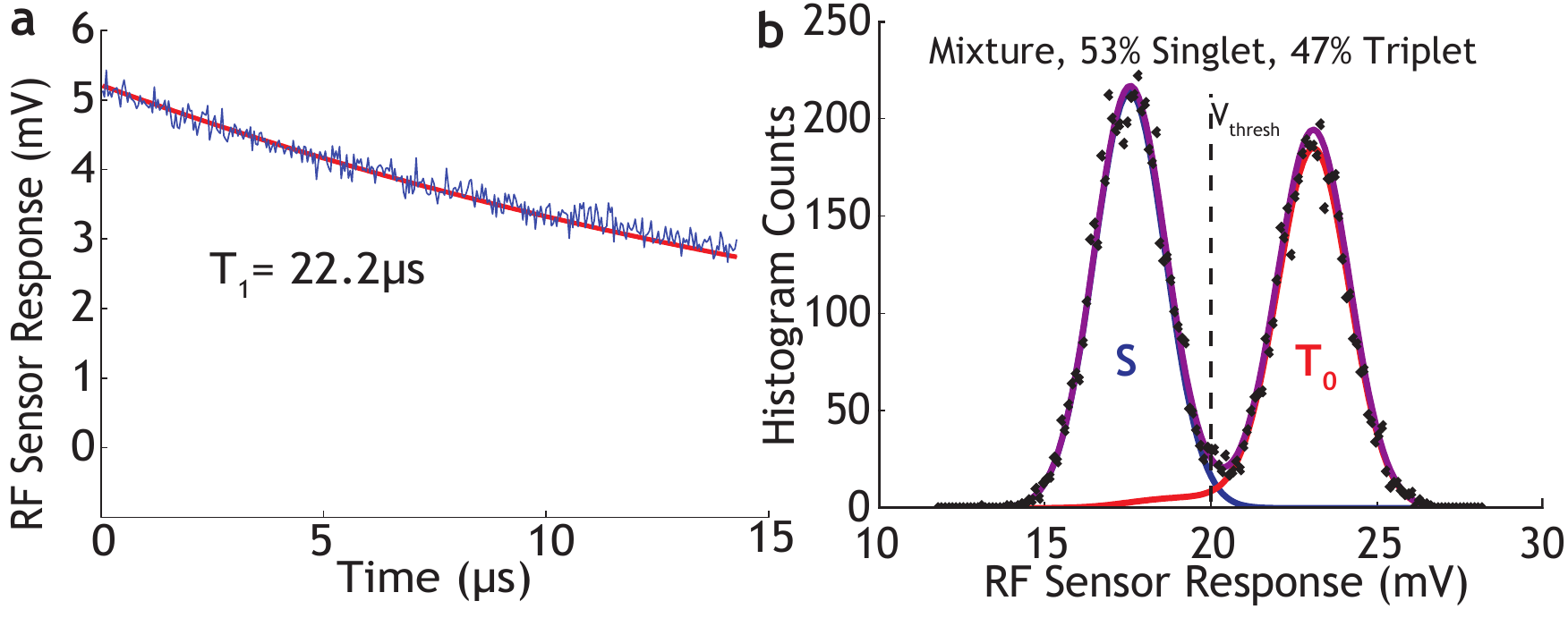}\caption{\textbf{Singleshot Readout: a,} The difference in sensor signal between
$|S\rangle$and $|T_{0}\rangle$ is fit to a decaying exponential
to determine $T_{1}$, which is used in calibration of sensor values.\textbf{
b,} The histograms of a mixture of $|S\rangle$ and $|T_{0}\rangle$
states used to calibrate the sensor values. If we choose a threshold
$V_{thresh}$ to distinguish between $|S\rangle$ and $|T_{0}\rangle$
we see a readout fidelity of 97\%. Purple: fit to noisy distribution
including $T_{1}$ decay from $|T_{0}\rangle$ to $|S\rangle$. The
deduced distribution for $|S\rangle$ (blue) is a Gaussian, while
that for $|T_{0}\rangle$ (red) has a tail due to $T_{1}$ decay. }
\end{figure}

\begin{figure*}
\includegraphics[scale=0.5]{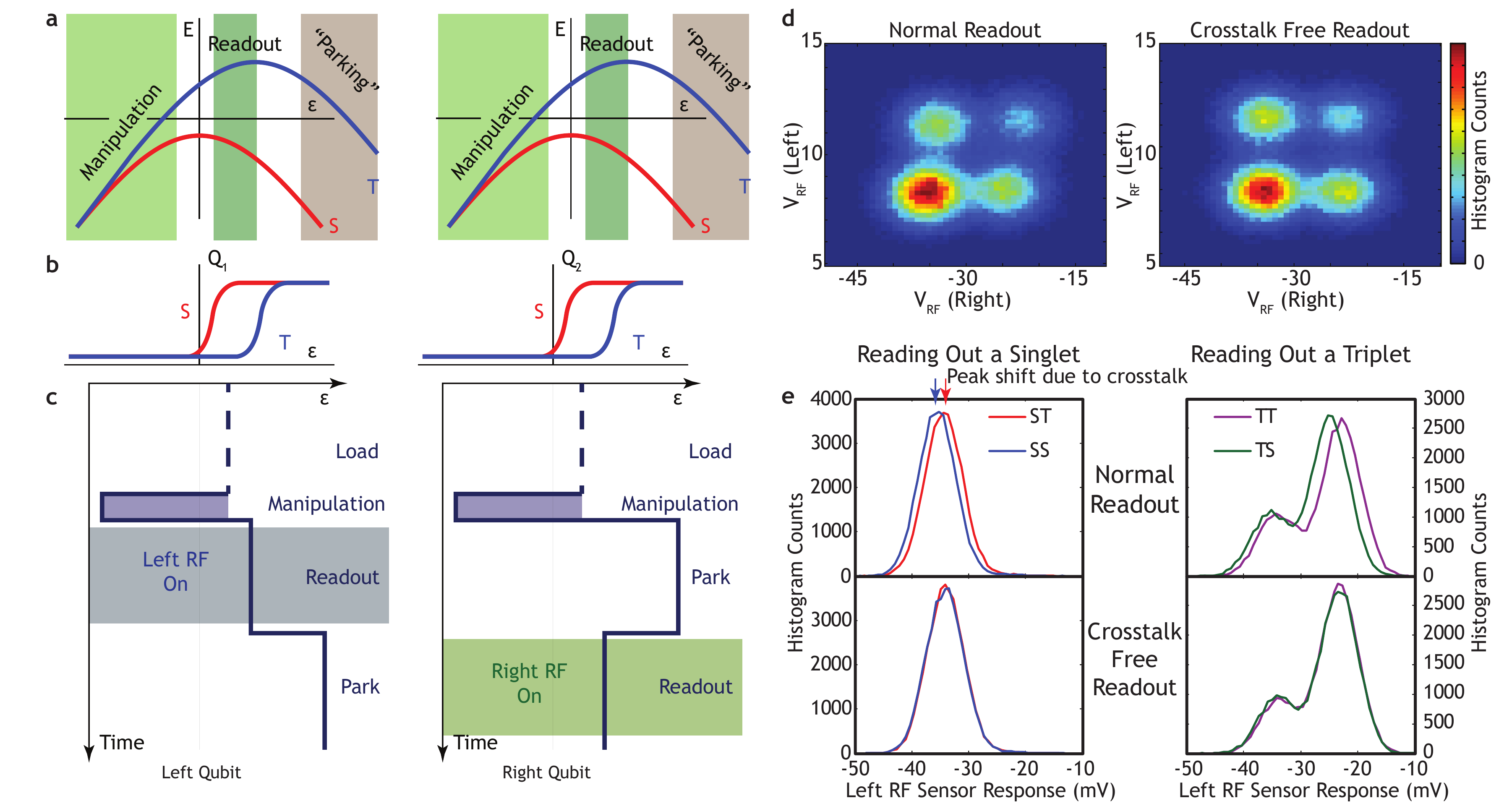}\caption{\textbf{Readout Crosstalk: a,} A schematic of the energy diagram as
a function of $\epsilon$ that describes the two qubits and shows
the regions of $\epsilon$ where different operations are carried
out. \textbf{b,} A schematic of the signal from the RF charge sensor
as a function of $\epsilon$ for the two qubits. This signal reflects
the charge distribution of the two qubit states. For large positive
$\epsilon$ there is a region where $|S\rangle$ and $|T_{0}\rangle$
have the same sensor signal (charge distribution), which is the foundation
of the crosstalk-free readout scheme. \textbf{c,} A schematic of the
readout scheme that eliminates crosstalk. First, the left qubit is
read while the right qubit is ``parked'' in (0,2), and then the
right qubit is read while the left qubit is ``parked'' in (0,2).
\textbf{d,} A two dimensional histogram of the RF sensor responses
without (left panel) and with (right panel) this crosstalk-free readout
scheme. \textbf{e,} Histograms of sensor values without (top) and
with (bottom) the crosstalk-free readout. Without the crosstalk-free
readout the sensor signal of one qubit depends on the state of the
other qubit. }
\end{figure*}

\textbf{\Large Readout Crosstalk}{\Large \par}

For accurate state tomography it is important that the readout of
the two qubits be independent. The two-qubit coupling relies on the
fact that the $|S\rangle$ and $|T_{0}\rangle$ in one qubit electrostatically
gate the other qubit, and reading out the two qubits simultaneously
leads to readout crosstalk; the left-qubit sensor value will be different
for $|S\rangle|T_{0}\rangle$ and $|S\rangle|S\rangle$, and similarly
for the right qubit (Suppl. Fig. 2d-e). We avoid this problem by reading
out the qubits sequentially: while one qubit is read out the other
qubit is ``parked'' deep inside of (0,2) where the (0,2)$|T_{0}\rangle$
is lower in energy than the (1,1)$|T_{0}\rangle$ so that both the
$|S\rangle$and $|T_{0}\rangle$ occupy (0,2) (Suppl. Fig. 2a-c).
In this way each qubit is read out while the other qubit has the same
charge distribution for the both qubit states. $T_{1}$ in this region
(with the RF excitation off) is large, so no measurable degradation
of the qubit being stored occurs.

\textbf{\Large Calibration of State Tomography}{\Large \par}

State tomography involves reading the projection of a qubit on to
the three Cartesian axes of the Bloch sphere\textsuperscript{15}.
However, the charge sensor next to each qubit allows us to determine
only the projection on to the z-axis (S-$T_{0}$) of the qubit. Therefore,
in order to perform state tomography, we apply rotations that map
the x and y axes of the qubit to the z-axis\textsuperscript{11}.
The y-axis component is mapped on to the minus z-axis by a $\pi/2$-pulse
around the x-axis, which is driven by $\Delta B_{z}$, and the x-axis
component is mapped on to the z-axis by adiabatically turning on $J$,
which maps the eigenstate $|\uparrow\downarrow\rangle$($|\downarrow\uparrow\rangle$)
into $|S\rangle$ ($|T_{0}\rangle$) (Fig. 2a). 

Extreme care must be taken when performing two-qubit state tomography,
as errors in tomography can introduce spurious correlations which
might inflate the measured degree of entanglement. However, in S-$T_{0}$
qubits, traditional methods for calibration are difficult to implement.
The two control axes, $J$ and $\Delta B_{z}$, are not orthogonal,
and precise 90 degree rotations are problematic due to the timing
resolution available on our signal generators as well as pulse rise
times. Pulse rise times in the experimental apparatus prevent instantaneous
changes in $J$, which changes both the axis of rotation and the total
angle rotated. These rise times also cause sudden changes in $J$
to become somewhat adiabatic. Furthermore, the effects of these pulse
rise times depend both on the starting and ending point of the an
individual pulse. Additionally, the two rotations involved in tomography
of S-$T_{0}$ qubits are not easily tuned: the accuracies of both
the x and y-axis readouts are limited by the speed of the waveform
generators ($\sim$1ns in this work). Therefore, we perform careful
measurements in order to determine the three axes on to which the
qubit is projected, and apply a transformation in order to map these
axes onto the traditional Cartesian axes. Dephasing during readout
rotations is represented in this process as readout axes that are
longer than the radius of Bloch sphere.

\begin{figure}
\includegraphics[scale=0.4]{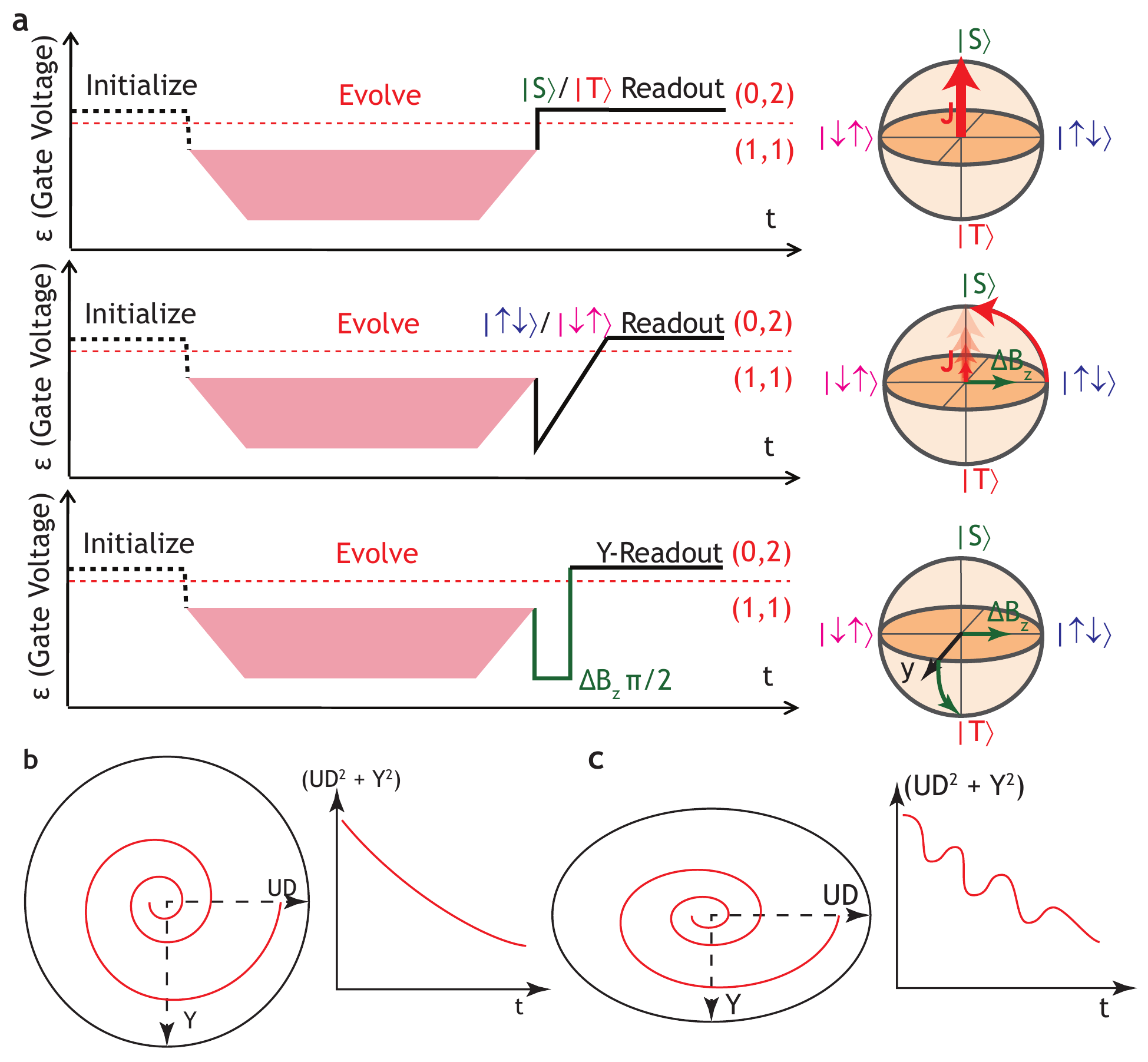}\caption{\textbf{Calibration of state tomography: a}, A schematic of how state
tomography is performed in S-$T_{0}$ qubits\textsuperscript{11}.
The S-$T_{0}$ component is read out by charge sensing in a region
where $|S\rangle$ occupies (0,2) and $|T_{0}\rangle$ occupies (1,1).
The x (y) component is read adiabatically turning on $J$ (rotating
by $\pi/2$ around the x-axis), followed by charge sensing. \textbf{b},
A schematic of the length of the Bloch vector for perfect state tomography.
\textbf{c}, A schematic of the ``ripples'' in the length of the
Bloch vector if state tomography is flawed. The tomographic axes are
determined by minimizing these ripples.}
\end{figure}

The procedure for determining the the tomographic axes makes minimal
assumptions. We first calibrate the the S-$T_{0}$ readout using the
singleshot histograms (see above). We then assume that any path of
the state of the qubit around the Bloch sphere in a free-induction-decay
experiment should smoothly dephase, i.e., there should not be oscillations
in the amplitude of the Bloch vector. No assumptions are made about
the axis or frequency of any rotation. If either the length or the
angle of one of the tomography axes is incorrect, we expect to see
``ripples'' in the length of the measured Bloch vector as a function
of evolution time(Fig. 3b-c). If we simultaneously consider many paths
of evolution around the Bloch sphere, errors in the angles and lengths
of the different readout axes become distinguishable due to varying
phases, periods, and amplitudes of the ripples in the lengths of the
measured vectors (Suppl. Fig. 4a). Therefore, to calibrate our axes,
we gather data on many different evolutions around the Bloch sphere
by evolving from many different starting points at many values of
$\epsilon$ (Suppl. Fig. 4b,c,e,f). We determine the axes on to which
we project our state by finding the axes that minimize the amplitude
of the ripples in the length of the Bloch vectors (Suppl. Fig. 4d).
Based on our measurement procedure, we define the S-$T{}_{0}$ axis
to lie along the z-axis. We allow the y-axis to lie anywhere on the
Bloch sphere because a rotation around the x-axis can suffer from
over/under rotation as well as adiabaticity issues with switching
$J$ on and off instantly. We constrain the x-axis to lie in the x-z-plane
because the only expected error is due to adiabaticity turning $J$
on and off. The typical tomographic axes are shown in Suppl. Fig.
4d, and the signs of the errors are consistent with their origins.
The variation from calibration to calibration is $\sim$1\% on the
axis lengths and angles. 

\begin{figure}[t]
\includegraphics[scale=0.5]{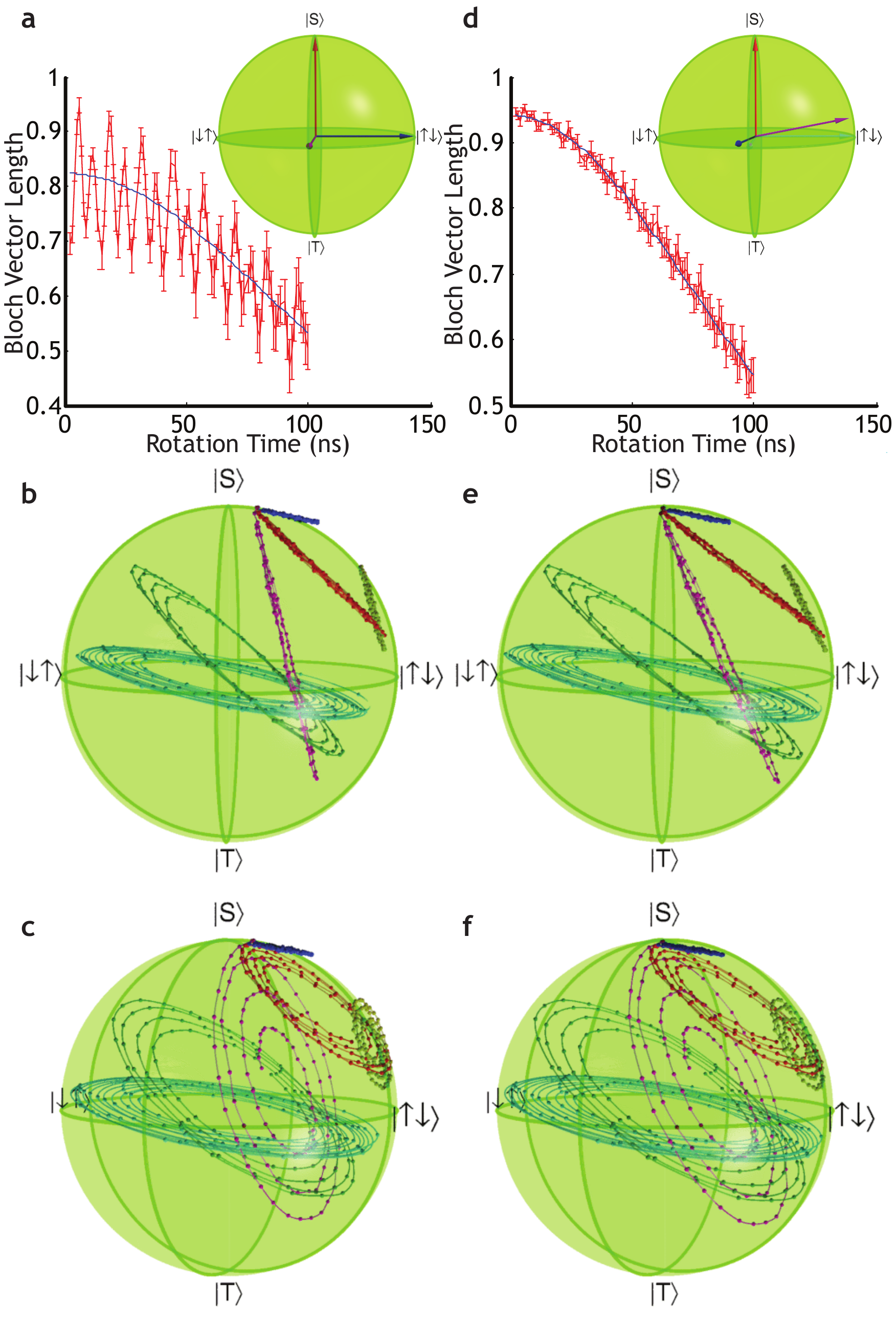}\caption{\textbf{Calibrated vs uncalibrated state tomography: a, }Data taken
to calibrate the tomography shows ripples in the length of the Bloch
vector if we assume that the tomography projects the quantum state
on to Cartesian axes (inset).\textbf{ b-c,} The paths around the Bloch
sphere for the different evolutions that are used for tomography calibration.
If the tomography is assumed to project on to the Cartesian axes there
are points that lay outside the Bloch sphere, and the pure $|S\rangle$
states are not at the north pole, which is indicative of flawed state
tomography. \textbf{d,} The ripples in the length of the bloch vector
are diminished (compared to panel a) if the axes deduced from state
tomography (inset) are used. \textbf{e-f,} The paths around the Bloch
sphere for the different evolutions that are used for state tomography.
When the correct axes are used, all the points lie inside the Bloch
sphere and the pure $|S\rangle$ are at the north pole. }
\end{figure}

\newpage
\textbf{\Large Determining Single Qubit Rotations}{\Large \par}

During the entangling sequence the two qubits rotate very rapidly
around the S-$T_{0}$ axis compared to the speed of the CPHASE gate
($\nicefrac{J_{1}}{2\pi}\sim\nicefrac{J_{2}}{2\pi}\sim300MHz$, $\nicefrac{J_{12}}{2\pi}\sim1MHz$).
These single qubit rotations are not perfectly canceled out by the
$\pi$-pulses in the dynamically decoupled sequence due to pulse distortions,
consistent with pulse rise time effects at short times and capacitive
coupling to RC-filtered DC gates at long times. Moreover, the angles
by which the qubits are rotated change as a function of the evolution
time $\tau$. In order to undo these rotations, we perform a least-square
fit of the data to the expected form of the Pauli set (see equation(\ref{eq:pauli set})
below), restricting the rotation to be around the S-$T_{0}$ axis
because $J_{1},J_{2}\gg\Delta B_{z}$. These angles are shown in Fig.
3b, and exhibit a smooth, monotonic behavior. The angles increase
quickly for small $\tau$, which is consistent with pulse rise time
effects, and display linear behavior for long $\tau$, which is consistent
with long time RC filtering. For comparison, we plot the entire Pauli
set for both the rotated and unrotated data in Suppl. Fig. 5 c-d.

\begin{figure}
\includegraphics[scale=0.5]{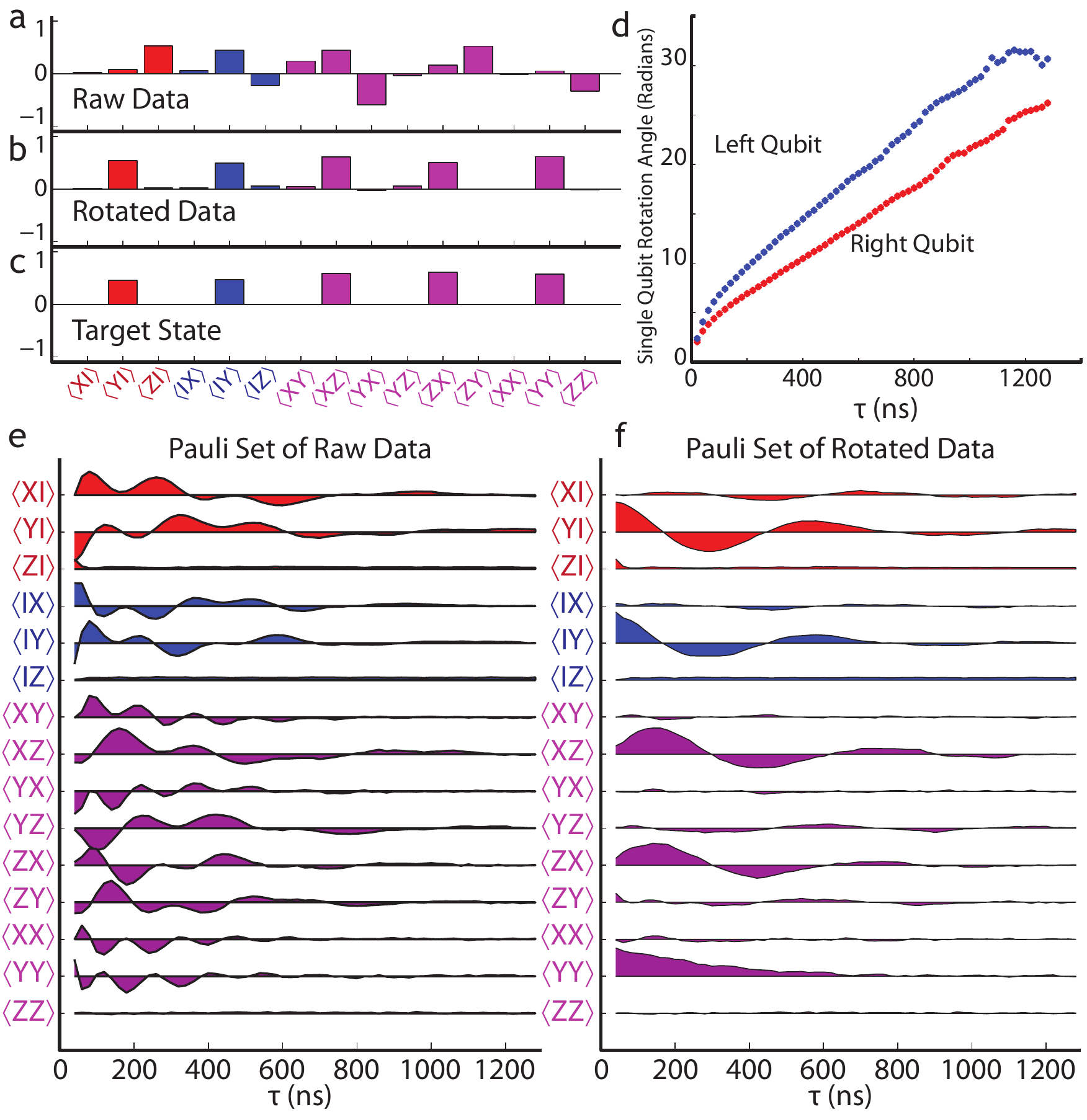}\caption{\textbf{Single-qubit rotations: a}, The Pauli set for $\tau=$100ns
as measured is complicated by single qubit rotations. \textbf{b,}
Numerically rotating each qubit around the S-$T_{0}$-axis simplifies
presentation and analysis. \textbf{c,} The expected state for $\tau=$100ns.
\textbf{d}, The single qubit rotation angles for both qubits as a
function of $\tau$ are smooth and monotonic functions. \textbf{e-f}
The entire Pauli set as a function of $\tau$ for the raw and rotated
data equation(\ref{eq:pauli set}). The the y-axes of adjacent elements
in the Pauli set are offset by 1. }
\end{figure}

\newpage
\textbf{\Large Fit Form for a Dephased Bell State}{\Large \par}

In order to fit the experimentally measured Pauli sets we calculate
$\rho\left(t\right)$ in the presence of fluctuations on $J_{1}$,
and $J_{2}$, and average over all these fluctuations\textsuperscript{25,26}.
For the present case where $J_{12}\ll J_{1},\, J_{2}$ , we neglect
the two-qubit dephasing, which is smaller than single-qubit dephasings
by a factor of $\frac{J_{1}}{J_{12}}$, $\frac{J_{2}}{J_{12}}\approx$
300. This yields the following non-zero elements of the Pauli set

\begin{eqnarray}
\langle YI\rangle & = & e^{-\nicefrac{\tau}{T_{2,1}}}\cos\left(J_{12}\tau\right)\nonumber \\
\langle IY\rangle & = & e^{-\nicefrac{\tau}{T_{2,2}}}\cos\left(J_{12}\tau\right)\nonumber \\
\langle XZ\rangle & = & e^{-\nicefrac{\tau}{T_{2,1}}}\sin\left(J_{12}\tau\right)\nonumber \\
\langle ZX\rangle & = & e^{-\nicefrac{\tau}{T_{2,2}}}\sin\left(J_{12}\tau\right)\nonumber \\
\langle YY\rangle & = & e^{-\nicefrac{\tau}{T_{2,1}}}e^{-\nicefrac{\tau}{T_{2,2}}}\label{eq:pauli set}
\end{eqnarray}

where $T_{2,i}$ are the single-qubit coherence times of the two qubits.
From fits of the data to this form we extract for $\delta\epsilon=$0:
$T_{2,1}=$420ns, $T_{2,2}=$510ns, and $\nicefrac{J_{12}}{2\pi}=$0.87MHz.
Imperfect state preparation, in which the $\pi/2$-pulse does not
leave the state of the qubits in the x-y-plane causes mixing between
terms, which is visible in the form of small amplitude oscillations
in the $\langle XI\rangle$,$\langle IX\rangle$, and $\langle XX\rangle$
components of the Pauli set (Suppl. Fig. 5f).